\documentclass[pra,twocolumn,superscriptaddress,showpacs]{revtex4}

\usepackage[dvips]{graphicx}
\usepackage{amsmath,amssymb}
\usepackage{bm}

\begin{document}
\title{Selective entanglement breaking}

\author{Yuma Kinoshita}
\email{kinosita@qi.mp.es.osaka-u.ac.jp}
\affiliation{Division of Materials Physics, 
Department of Materials Engineering Science,
Graduate school of Engineering Science, Osaka University, 
1-3 Machikaneyama, Toyonaka, Osaka 560-8531, Japan}
\affiliation{CREST Reserch Team for Photonic Quantum Information,
4-1-8 Honmachi, Kawaguchi, Saitama 331-0012, Japan}

\author{Ryo Namiki}
\affiliation{Division of Materials Physics, 
Department of Materials Engineering Science,
Graduate school of Engineering Science, Osaka University, 
1-3 Machikaneyama, Toyonaka, Osaka 560-8531, Japan}
\affiliation{CREST Reserch Team for Photonic Quantum Information,
4-1-8 Honmachi, Kawaguchi, Saitama 331-0012, Japan}

\author{Takashi Yamamoto}
\affiliation{Division of Materials Physics, 
Department of Materials Engineering Science,
Graduate school of Engineering Science, Osaka University, 
1-3 Machikaneyama, Toyonaka, Osaka 560-8531, Japan}
\affiliation{CREST Reserch Team for Photonic Quantum Information,
4-1-8 Honmachi, Kawaguchi, Saitama 331-0012, Japan}

\author{Masato Koashi}
\affiliation{Division of Materials Physics, 
Department of Materials Engineering Science,
Graduate school of Engineering Science, Osaka University, 
1-3 Machikaneyama, Toyonaka, Osaka 560-8531, Japan}
\affiliation{CREST Reserch Team for Photonic Quantum Information,
4-1-8 Honmachi, Kawaguchi, Saitama 331-0012, Japan}
\affiliation{SORST Reserch Team for Interacting Carrier Electronics,
4-1-8 Honmachi, Kawaguchi, Saitama 331-0012, Japan}

\author{Nobuyuki Imoto}
\affiliation{Division of Materials Physics, 
Department of Materials Engineering Science,
Graduate school of Engineering Science, Osaka University, 
1-3 Machikaneyama, Toyonaka, Osaka 560-8531, Japan}
\affiliation{CREST Reserch Team for Photonic Quantum Information,
4-1-8 Honmachi, Kawaguchi, Saitama 331-0012, Japan}
\affiliation{SORST Reserch Team for Interacting Carrier Electronics,
4-1-8 Honmachi, Kawaguchi, Saitama 331-0012, Japan}

\date{\today}

\begin{abstract}
We discuss the cases where local decoherence selectively degrades one
type of entanglement more than other types. A typical case is called 
state ordering change, in which two input states with different amounts 
of entanglement undergoes a local decoherence and the state with the 
larger entanglement results in an output state with less entanglement 
than the other output state. We are also interested in a special case
where the state with the larger entanglement evolves to a separable 
state while the other output state is still entangled, which we call 
selective entanglement breaking. 
For three-level or larger systems,
it is easy to find examples of the state ordering change and 
the selective entanglement breaking,
but for two-level systems it is not trivial whether such situations exist.
We present a new strategy to construct examples of two-qubit states
exhibiting the selective entanglement breaking
regardless of entanglement measure.
We also give a more striking example of the selective entanglement breaking
in which the less entangled input state
has only an infinitesimal amount of entanglement. 
\end{abstract}

\maketitle
\section{Introduction}
Quantum entangled states on a composite system 
are vital resources for many quantum information
protocols \cite{1}, and it is important to understand how various
entangled states are affected by decoherence when one of the local
subsystems is interacted with the environment or is 
transferred over a noisy quantum channel. 
Here we focus on the cases in which the effects of such 
a local quantum operation/channel 
are different on two different kinds of entanglement.
In a typical situation,
one input state has a large amount of entanglement and 
the other one has very small entanglement, but after applying a local
quantum channel on one of the 
systems, the entanglement in the former state is completely destroyed
while the latter state is still entangled. In other words, the less
entangled state is robust against the noises of the quantum channel 
that severely degrades the other type of entanglement.

When the dimension of Hilbert space for the local system is more than
two, namely, for a three-level or larger system, 
such an example is easy to find \cite{15}. 
Since two levels (a two-dimensional subspace)
are enough to form entanglement to another system, and there are 
different pairs of levels to choose, it is easy to imagine 
a quantum channel which completely destroys the coherence between 
a specific pair of levels, while leaving the coherence between another 
pair intact. If, however, the system in question is a qubit (a two-level system),
the problem becomes nontrivial
because any entanglement must use the whole two-dimensional space.
Hence the phenomenon of ``state ordering change'' by 
local evolution has been sought after 
\cite{15}, in which the amount of entanglement $E$ satisfies
\begin{eqnarray}
\displaystyle
E(\hat{\omega}_1) \ge E(\hat{\omega}_2), \;\; E({{\hat{\omega}}'}_1) \le E({{\hat{\omega}}'}_2)
\label{eq:changeorderequal}
\end{eqnarray}
for input states $\hat{\omega}_1, \hat{\omega}_2$ and the output states
${\hat{\omega}'}_j=\mathcal{E}_L [\hat{\omega}_j] \ (j=1,2)$, where 
$\mathcal{E}_L = \mathcal{E} \otimes \mathcal{I}$ is a local quantum channel.
Ziman and Bu$\check{\rm z}$ek have found examples of such state ordering change 
for a particular measure $E$ of entanglement \cite{15}.

At this point, one must recall that the ordering between two entangled
states may depend on the choice of the entanglement measure. 
Eisert and Plenio showed that the condition, $E'(\hat{\omega}_1) <
E'(\hat{\omega}_2) \Leftrightarrow E''(\hat{\omega}_1) <
E''(\hat{\omega}_2)$, is not always satisfied from Monte Carlo
simulation ($E', E''$ are two different entanglement measures)
\cite{11}. That is to say, there exists a pair of states with 
$E'(\hat{\omega}_1) < E'(\hat{\omega}_2)$ and $E''(\hat{\omega}_1) >
E''(\hat{\omega}_2)$. 
Miranowicz and Grudka studied such an ambiguity in the ordering in
two-qubit states for  entanglement measures including negativity,
concurrence, and relative entropy of entanglement \cite{12,13}. 
Hence a measure-dependent example is not enough to ascertain 
the existence of the state ordering change in a local qubit channel.
For a definite answer, we need to show that Eq.~(\ref{eq:changeorderequal})
holds for any measure of entanglement $E$.

In this paper, we propose a general strategy to produce
many examples of two-qubit states and a qubit channel showing 
the state ordering change for any measure of entanglement.
In these examples, the output state 
$\hat{\omega}'_1=\mathcal{E}_L [\hat{\omega}_1]$ is separable,
namely, the qubit channel destroys the entanglement in state $\hat{\omega}_1$
completely but leaves behind part of the entanglement in state
$\hat{\omega}_2$,
which we call ``selective entanglement breaking''.
We further show that a particular example constructed from 
the above strategy exhibits a more striking feature that the
channel breaks entanglement in $\hat{\omega}_1$ selectively 
even when the input state 
$\hat{\omega}_2$ has an infinitesimal amount of entanglement.

The construction of the paper is as follows.
In Sec. \ref{sec:qutrit},
we give a trivial example of two-qutrit states and a qutrit
channel showing the selective entanglement breaking.
We also give precise definitions of the three relevant 
phenomena: state ordering change,
selective entanglement breaking, and 
strong selective entanglement breaking. 
In the main part of the paper, Sec. \ref{sec:qubit},
we present a strategy for finding
examples of two-qubit states and a qubit channel showing
the selective entanglement breaking.
We also construct a specific example and 
show that it exhibits the strong selective entanglement breaking.
In Sec. \ref{sec:pure},
we show there is no selective entanglement breaking
for two-qubit pure states.
In Sec. \ref{sec:4-strict-state}, we consider a family of entanglement measures
for which the state ordering change can be discussed with a strict inequality.
Finally Sec. \ref{sec:conclu} concludes the paper.
 
\section{Two-qutrit state ordering change}
\label{sec:qutrit}

Let us consider the case where the local system is a qutrit, namely, 
the dimension of the Hilbert space is three. We can easily find 
an example of two-qutrit states and a qutrit channel showing the state
ordering change. Consider two pure states $|{\psi_1}\rangle = \sqrt{1/2}
(|{11}\rangle + |{22}\rangle )$ and $|{\psi_2}\rangle =
\sqrt{q}|{00}\rangle + \sqrt{1-q}|{11}\rangle$ ($0 < q < 1$). 
Suppose that the local channel $\mathcal{E}$ applied to the first system 
is represented by Kraus operators (the operator-sum representation):
$\mathcal{E}(\hat{\rho})=\sum_j \hat{M}_j \hat{\rho} \hat{M}_j^\dagger$,
where $\hat{M}_0=|{0}\rangle \langle{0}|+|{1}\rangle \langle{1}|$ and
$\hat{M}_1=|{2}\rangle \langle{2}|$.
After applying $\mathcal{E}_L = \mathcal{E}
\otimes \mathcal{I}$ to the two input states, we obtain 
a separable state $\hat{\rho}^{\rm{out}}_1=(|{11}\rangle
\langle{11}|+|{22}\rangle \langle{22}|)/2$ for the first input state, 
but the second input state remains unaltered, 
$\hat{\rho}^{\rm{out}}_2 =|{\psi_2}\rangle\langle {\psi_2}|$.

In this example, the input state $|{\psi_1}\rangle$
can be transformed to $|{\psi_2}\rangle$ by
local operations and classical communication (LOCC).
One of the parties applies a local filter described by Kraus 
operators
$\hat{A}_0=|{0}\rangle \langle{0}|+\sqrt{1-q}|{1}\rangle
\langle{1}|+\sqrt{q}|{2}\rangle \langle{2}|$ and  
$\hat{A}_1=\sqrt{q}|{1}\rangle \langle{1}|+\sqrt{1-q}|{2}\rangle
\langle{2}|$, and classically communicate the outcome (0 or 1)
to the other party. Then, local unitary operations can transform the 
filtered states as
$\sqrt{1-q}|{11}\rangle + \sqrt{q}|{22}\rangle {\longrightarrow}
\sqrt{q}|{00}\rangle + \sqrt{1-q}|{11}\rangle$ and
$\sqrt{q}|{11}\rangle+\sqrt{1-q}|{22}\rangle {\longrightarrow}
\sqrt{q}|{00}\rangle + \sqrt{1-q}|{11}\rangle$. 
We can also transform $\hat{\rho}^{\rm{out}}_2$ to 
$\hat{\rho}^{\rm{out}}_1$ by LOCC, since the latter is separable. 
These observations assure that we have $E(\psi_1)\ge E(\psi_2)$
and $E(\hat{\rho}^{\rm{out}}_1) \le E(\hat{\rho}^{\rm{out}}_2)$
for any entanglement measure $E$ as long as it satisfies \\

(i)\textit{Monotonicity under LOCC}, LOCC cannot increase the
entanglement, namely, if the state $\hat{\rho}^{AB}$ 
is transformed into $\hat{\sigma}^{AB}$ by LOCC, 
$E(\hat{\rho}^{AB}) \ge E(\hat{\sigma}^{AB})$.\\

\noindent
We see that this trivial example shows the state ordering change
regardless of the choice of entanglement measure,
which we define formally as \\

{\em State ordering change} --- A local quantum channel 
$\mathcal{E}_L = \mathcal{E} \otimes \mathcal{I}$ 
and two input states $\hat{\omega}_1$
and $\hat{\omega}_2$ satisfy
\begin{eqnarray}
E(\hat{\omega}_1) \ge E(\hat{\omega}_2), \;\; 
E(\mathcal{E}_L [\hat{\omega}_1]) \le 
E(\mathcal{E}_L [\hat{\omega}_2])
\label{eq:withequality}
\end{eqnarray}
for any entanglement measure $E$ satisfying the monotonicity 
under LOCC, and among such measures, there exists 
a measure $E'$ satisfying
\begin{eqnarray}
E'(\hat{\omega}_1) > E'(\hat{\omega}_2), \;\; 
E'(\mathcal{E}_L [\hat{\omega}_1]) < 
E'(\mathcal{E}_L [\hat{\omega}_2]).
\label{eq:strict}
\end{eqnarray}

Here we cannot demand the strict inequality to hold for 
any measure, since the property (i) cannot exclude a 
trivial measure which is constant for any state whether 
it is entangled or not. If we are to require the 
strict inequality, we need to restrict the allowed 
entanglement measures, which will be discussed in 
Sec. \ref{sec:4-strict-state}.

In the above example of two-qutrit state ordering change, 
the entanglement in the state $\omega_1$ is completely 
destroyed by the channel.
In this paper, we define such cases as follows:\\

{\em Selective entanglement breaking} ---
The state ordering change occurs with one of the output states 
being separable.

Moreover, in the trivial example considered here, the entanglement 
in state $|\psi_2\rangle$ is preserved no matter how small 
its entanglement is. This implies that the quantum channel 
selectively destroys the type of entanglement held 
in a state $\hat{\omega}_1$, while it does not completely destroy
entanglement held in another state $\hat{\omega}_2$ even when
its entanglement of formation \cite{20},
$E_f(\hat{\omega}_2)$, is infinitesimal.
Here we will define such a phenomenon in the following way.\\

{\em Strong selective entanglement breaking} --- 
A local quantum channel 
$\mathcal{E}_L = \mathcal{E} \otimes \mathcal{I}$, 
an input state $\hat{\omega}_1$,
and a sequence of input states $\{\hat{\omega}_j\}_{j=2,3\ldots
\infty}$ satisfy (a) $E(\hat{\omega}_1) \ge E(\hat{\omega}_2) \ge 
E(\hat{\omega}_3) \ge \cdots$ for any measure $E$ 
satisfying (i), (b) There is a measure $E'$ such that 
$E'(\hat{\omega}_1) > E'(\hat{\omega}_2)$ and 
$E'(\mathcal{E}_L [\hat{\omega}_1]) < 
E'(\mathcal{E}_L [\hat{\omega}_j])$
for any $j\ge 2$, (c)
$\lim_{j \to \infty} {E_f}(\hat{\omega}_j)=0$,
and (d) $\mathcal{E}_L [\hat{\omega}_1]$ is separable.

In the three-level system 
considered here
or in larger systems, the existence of 
the state ordering change is trivial, and even the existence of 
the strong selective entanglement breaking
is also trivial as shown above.
The next section will deal with the nontrivial question 
for the case of a two-level system.

\section{Two-qubit state ordering change}
\label{sec:qubit}

In this section, we show an example of the selective entanglement breaking
where the local system is a qubit.
For two-qutrit system, it was not difficult to find the example of
the strong selective entanglement breaking. This is because
for three-level or larger systems we can preserve entanglement
even if we apply a projector onto a two-dimensional subspace. 
But for a two-level input system, we cannot preserve entanglement by 
nontrivial projections. It is thus difficult to find an example of 
two-qubit selective entanglement breaking along the line used in the previous section.

Our strategy to find such an example is as follows. 
First we consider an entangled mixed state $\hat{\rho}^{AB}$ 
and a qubit channel $\mathcal{E}_\lambda$ with 
a parameter $\lambda$ representing the amount of the noise introduced by the 
channel. After applying the local channel to the input state 
$\hat{\rho}^{AB}$, we calculate the negativity \cite{4,5} 
of state $\mathcal{E}_\lambda \otimes \mathcal{I}(\hat{\rho}^{AB})$
as a function of $\lambda$ to find the value of $\lambda=\lambda_{\rm sep}$ at which the state 
becomes separable. Next, going back to the original state
$\hat{\rho}^{AB}$,
we apply a local unitary $\hat{U}$
to the first system to produce $\hat{\rho}^{AB}_U\equiv 
(\hat{U}\otimes \hat{I}) \hat{\rho}^{AB}(\hat{U}^\dagger\otimes \hat{I}^\dagger)$.
We again calculate the critical value $\lambda_{\rm sep}$ for this state.
The success of our strategy rests on whether the critical value $\lambda_{\rm sep}$
changes depending on the choice of the unitary $\hat{U}$.
Once we find such a dependency, we can construct an example of the selective entanglement breaking
as follows. Without loss of generality,
we can assume that there is a value of $\lambda$ for which 
$\mathcal{E}_\lambda \otimes \mathcal{I}(\hat{\rho}^{AB}_U)$ is separable, 
while the negativity of state 
$\mathcal{E}_\lambda \otimes \mathcal{I}(\hat{\rho}^{AB})$ is strictly 
positive. Then, we consider an LOCC operation $\mathcal{E}'_\epsilon$
with parameter $\epsilon$ representing the strength of the noise
($\mathcal{E}'_0=\mathcal{I}\otimes \mathcal{I}$), and apply it to state
$\hat{\rho}^{AB}$ to obtain $\hat{\rho}^{AB}_{\epsilon}\equiv
\mathcal{E}'_\epsilon(\hat{\rho}^{AB})$.
If $\epsilon$ is small enough but nonzero, we have
$N(\hat{\rho}^{AB}_{\epsilon})< N (\hat{\rho}^{AB})=N (\hat{\rho}^{AB}_U)$
while the negativity of the new state
$\mathcal{E}_\lambda \otimes \mathcal{I}(\hat{\rho}^{AB}_{\epsilon})$
should still be strictly positive, namely,
$N(\mathcal{E}_\lambda \otimes \mathcal{I}(\hat{\rho}^{AB}_{\epsilon}))
>0=N(\mathcal{E}_\lambda \otimes \mathcal{I}(\hat{\rho}^{AB}_U))$.
Hence, for the negativity, the strict inequality (\ref{eq:strict})
is satisfied (see Fig.1). On the other hand, we can convert $\hat{\rho}^{AB}_U$
to $\hat{\rho}^{AB}_{\epsilon}$ by LOCC ($\hat{U}^{-1}\otimes \hat{I}$
followed by $\mathcal{E}'_\epsilon $), and 
we can also convert $\mathcal{E}_\lambda \otimes
\mathcal{I}(\hat{\rho}^{AB}_{\epsilon})$ to 
$\mathcal{E}_\lambda \otimes \mathcal{I}(\hat{\rho}^{AB}_U)$
by LOCC since the latter is separable. Hence, from the 
monotonicity (i), the inequality (\ref{eq:withequality})
holds for any measure $E$.
\begin{figure}[t]
\begin{center}
\includegraphics[width=8.6cm]{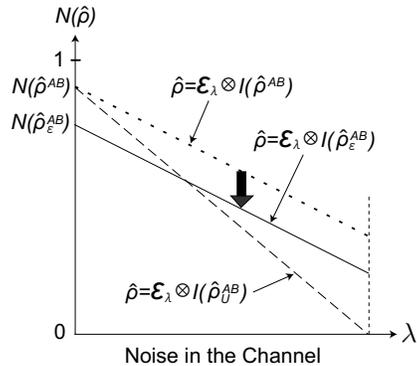}
\caption{Strategy to find examples of the selective entanglement breaking. 
We use negativity $N(\hat{\rho})$ as an entanglement measure,
since it is easy to calculate.
Slight decrease in the degree of entanglement (the arrow) is 
done by LOCC, and hence the order of the two input states is 
measure-independent. The order of the two output states is 
also measure-independent, since one of them is separable.
\label{fig:1}}
\end{center}
\end{figure}

Let us show a specific example using the above strategy. 
First we consider a mixed entangled state, 
\begin{eqnarray}
\displaystyle
\hat{\rho}_{1}^{\rm in}=\frac{2}{3}|{\Phi_+}\rangle  \langle{\Phi_+}|+\frac{1}{3}|{00}\rangle  \langle{00}|,
\label{eq:rho1rho2}
\end{eqnarray}
where $|{\Phi_+}\rangle=\sqrt{1/2}(|{00}\rangle + |{11}\rangle) $. 
Consider a local unitary on the first system $A$,
\begin{eqnarray}
\displaystyle
\hat{U}_A=
\frac{1}{\sqrt{2}}
\left(
\begin{array}{cc}
1&-1 \\
1&1 \\
\end{array} 
\right),
\label{eq:unitary}
\end{eqnarray}
which is the rotation around $Y$ axis by $\theta =\pi/2$ on the Bloch sphere.
Let $\hat{\rho}_2^{\rm{in}}=(\hat{U}_{A} \otimes
\hat{I}_{B})\hat{\rho}^{\rm {in}}_1(\hat{U}_{A}^\dagger \otimes
\hat{I}_{B}^\dagger)$, which is written on the 
basis $\{|00\rangle,|01\rangle,|10\rangle,|11\rangle\}$ in the matrix form
\begin{eqnarray}
\displaystyle
\hat{\rho}_2^{\rm{in}}=\frac{1}{6}
\left(
\begin{array}{cccc}
2& -1 & 2 & 1 \\
-1 & 1 & -1 & -1 \\
2 & -1 & 2 & 1 \\
1 & -1 & 1 & 1 \\
\end{array} 
\right). 
\label{eq:rho2in}
\end{eqnarray}

As the local quantum channel applied to system $A$, we take 
a phase damping channel represented by two Kraus operators
\begin{eqnarray}
\displaystyle
\hat{E}_0
=
\left(
\begin{array}{cccc}
1&0 \\
0&\sqrt{1-\lambda } \\
\end{array} 
\right),
\label{eq:e0}
\ \ 
\hat{E}_1
=
\left(
\begin{array}{cccc}
0&0 \\
0&\sqrt{\lambda } \\
\end{array} 
\right),
\label{eq:e1}
\end{eqnarray}
where $0\le \lambda \le 1$. 
For an input two-qubit state $\hat{\rho}^{\rm{in}}$,
the output of the phase damping channel is given by
$\hat{\rho}^{\rm{out}}=(\hat{E}_0 \otimes \hat{I}) \hat{\rho}^{\rm{in}}
(\hat{E}_0^\dagger \otimes \hat{I}^\dagger) + (\hat{E}_1 \otimes
\hat{I}) \hat{\rho}^{\rm{in}} (\hat{E}_1^\dagger \otimes
\hat{I}^\dagger)$. 
For the two input states $\hat{\rho}^{\rm{in}}_1$ and $\hat{\rho}^{\rm{in}}_2$, the output states are calculated as 
\begin{eqnarray}
\displaystyle
\hat{\rho}_1^{\rm{out}}&=& \frac{1}{3}
\left(
\begin{array}{cccc}
2& 0& 0& \sqrt{1-\lambda } \\
0& 0& 0& 0 \\
0& 0& 0& 0 \\
\sqrt{1-\lambda }& 0& 0& 1 \\
\end{array} 
\right),
\label{eq:rho1out}\\
\hat{\rho}_2^{\rm{out}}&=& \frac{1}{6}
\left(
\begin{array}{cccc}
2&-1&2\sqrt{1-\lambda }&\sqrt{1-\lambda } \\
-1&1&-\sqrt{1-\lambda }&-\sqrt{1-\lambda } \\
2\sqrt{1-\lambda }&-\sqrt{1-\lambda }&2&1 \\
\sqrt{1-\lambda }&-\sqrt{1-\lambda }&1&1 \\
\end{array} 
\right).\nonumber\\
\label{eq:rho2out}
\end{eqnarray}

The negativity for a bipartite state $\hat{\rho}$ is defined by
\begin{eqnarray}
\displaystyle
N(\hat{\rho})=\max{\left\{0,-2\mu_{\rm{min}}\right\}},
\label{eq:negativity}
\end{eqnarray}
where $\mu_{\rm{min}}$ is the minimum of the eigenvalues of the partial
transpose of state $\hat{\rho}$ \cite{4,5}. 
For two-qubit states, it has the range from 0 (separable) to 1 (maximally entangled).
The negativity of the two output states $\hat{\rho}_1^{\rm{out}}$ and $\hat{\rho}_2^{\rm{out}}$ are
\begin{figure}[t]
\begin{center}
\includegraphics[width=8.6cm]{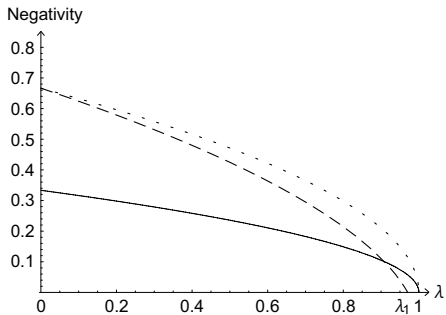}
\caption{Example of the selective entanglement breaking. The negativities of
$\hat{\rho}_1^{\rm{out}}$, $\hat{\rho}_2^{\rm{out}}$,
$\hat{\rho}_3^{\rm{out}}$ ($t=1/3$) are indicated by the dotted
curve, the dashed curve, and the solid curve, respectively. 
\label{fig:2}}
\end{center}
\end{figure}
\begin{eqnarray}
\displaystyle
N(\hat{\rho}_1^{\rm{out}})&=&\frac{2}{3}\sqrt{1-\lambda },\\
\label{eq:nega1out}
N(\hat{\rho}_2^{\rm{out}})&=&\frac{1}{6}\Bigl(-3+3\sqrt{1-\lambda }+\sqrt{10+6\sqrt{1-\lambda }-5\lambda }\Bigr),\nonumber\\
\label{eq:nega2out}
\end{eqnarray}
which are shown in Fig.~2.

When the noise parameter of the quantum channel is 
$\lambda=\lambda_1 \equiv-16+12\sqrt{2}\approx 0.97$,
$N(\hat{\rho}_2^{\rm{out}})=0$, and
$N(\hat{\rho}_1^{\rm{out}})_{|\lambda =\lambda_1} =2(3-2\sqrt{2})/3\approx 0.11>0$.
Then we decrease $N(\hat{\rho}_1^{\rm{in}})$ using LOCC. 
Specifically, we consider an operation in which 
the state is replaced by $|{00}\rangle$
with probability $\epsilon$.
Applying this operation to $\hat{\rho}_1^{\rm{in}}$, we have
\begin{eqnarray}
\displaystyle
\hat{\rho}_3^{\rm{in}}=t|{\Phi_+}\rangle  \langle{\Phi_+}|+(1-t)|{00}\rangle  \langle{00}|,
\label{eq:rho3out}
\end{eqnarray}
where $t=2(1-\epsilon)/3<2/3$. 
Calculating in a similar way, we obtain the negativity of the output
state as
\begin{eqnarray}
N(\hat{\rho}_3^{\rm{out}})=t \sqrt{1-\lambda }.
\label{eq:negarho3out}
\end{eqnarray}
The case with $t=1/3$ is shown 
in Fig.~2. 

As explained for the general strategy, 
$E(\hat{\rho}_3^{\rm{in}}) \le E(\hat{\rho}_2^{\rm{in}})$
and $E(\hat{\rho}_3^{\rm{out}}) \ge E(\hat{\rho}_2^{\rm{out}})$
hold for any entanglement measure satisfying the monotonicity (i).
On the other hand, $N(\hat{\rho}_3^{\rm{in}}) <
N(\hat{\rho}_2^{\rm{in}})$ holds for $t<2/3$ and 
$N(\hat{\rho}_3^{\rm{out}}) > N(\hat{\rho}_2^{\rm{out}})$
holds for $t>0$. Hence the phase damping channel with 
$\lambda=\lambda_1$ and the states $\hat{\rho}_2^{\rm{in}}$ and 
$\hat{\rho}_3^{\rm{in}}$ with $0<t<2/3$ exhibit the selective entanglement breaking.

In the limit of $t\rightarrow 0$, the state $\hat{\rho}_3^{\rm{in}}$
becomes separable. But as long as $t>0$, the output state is 
still entangled, and the state ordering change occurs. Hence the
particular example here exhibits not only the selective entanglement breaking, but
also the strong selective entanglement breaking.

\section{No selective entanglement breaking exists for two-qubit pure states}
\label{sec:pure}

In the previous section, we present a specific example of
selective entanglement breaking in which 
the two input states are mixed states.
In fact, we cannot find pure-state examples by our strategy.
The reason is closely related to the following general property of 
qubit channels: 
if a two-qubit pure entangled state 
becomes separable after one of the qubit passes through 
a qubit channel, then the channel is an
entanglement breaking channel \cite{16,17}, namely,
the channel destroys the entanglement of any input state.
We can prove it as follows.

Consider a two-qubit pure state
$|{\psi_{\rm{in}}} \rangle = \alpha |{00}\rangle + \beta |{11}\rangle$,
where ${\alpha}^2 +{\beta}^2=1$, and suppose that
$|{\psi_{\rm{in}}} \rangle$ becomes separable state $\hat\rho_{\rm sep}$
after the application of a qubit channel. Imagine we further 
apply a local filter, which is described by the Kraus operator
$\hat{M}_0=\beta|{0}\rangle \langle{0}|+\alpha |{1}\rangle \langle{1}|$,
to the second qubit. The state after the successful filtering,
$\hat\sigma_{\rm sep}\equiv (\hat{I}\otimes \hat{M}_0)\hat\rho_{\rm sep}
(\hat{I}\otimes \hat{M}_0)^\dagger$, is separable since $\hat\rho_{\rm sep}$
is separable. Now notice that 
even if we apply the local filter first and then apply the one-qubit channel,
the final state should be the same 
because they are operations on different systems.
In this case, the state after the successful filtering is 
a maximally entangled state
$\sqrt{1/2}(|{00}\rangle + |{11}\rangle)$, which 
evolves into the separable state $\hat\sigma_{\rm sep}$
after the application of the qubit channel. Hence the channel 
must be an entanglement breaking channel.

This property of qubit channels immediately tells us that 
there is no selective entanglement breaking with 
the input two-qubit state $\hat{\omega}_1$, which 
is to be broken, being a pure state. 
Our strategy in Sec.~\ref{sec:qubit} does not work for 
pure $\hat\rho^{AB}$ because
$\lambda_{\rm sep}$ is the same for 
any unitary operation $\hat{U}$. 
Although there is no selective entanglement breaking 
for pure two-qubit states, the present argument does not 
exclude the possibility of the state ordering change 
for two pure two-qubit input states.

For a system with dimension $d$ larger than two, a straightforward
extension of 
the above proof shows that a local channel is entanglement breaking 
if a pure entangled state {\em with full local rank} 
(the marginal density operator having rank $d$) is broken by the 
channel. This leaves the possibility of having the 
selective entanglement breaking of pure states with a small
local rank, as in the trivial example shown in Sec.~\ref{sec:qutrit}.
 
\section{State ordering change with strict inequality}
\label{sec:4-strict-state}

As discussed in Sec. \ref{sec:qutrit}, not all of the entanglement measures satisfying 
the monotonicity (i) fulfill the strict inequality (\ref{eq:strict}).
Here we show that, in the example in Sec.~\ref{sec:qubit}, 
the strict inequality (\ref{eq:strict}) holds for a wide range of 
measures specified by a set of additional conditions which are often
considered to be desirable as a measure of entanglement.
We consider the measures satisfying the following properties:\\
\\
(i$'$)\textit{Monotonicity under LOCC on average}, if LOCC transforms
$\hat{\rho}^{AB}$ into a state $\hat{\rho}^{AB}_i$ with probability
$p_i$, 
the entanglement does not increase on average, i.e.
$E(\hat{\rho}^{AB}) \ge \sum _i p_i E(\hat{\rho}^{AB}_i)$;\\
(ii)\textit{Vanishing on separable states}, $E(\hat{\rho}^{AB})=0 $ if $\hat{\rho}^{AB}$ is separable;\\
(iii)\textit{Normalization}, $E(|\Phi \rangle \langle \Phi | )=1$, where $|\Phi \rangle=\sqrt{1/2}(|{00}\rangle + |{11}\rangle)$;\\
(iv)\textit{Convexity}, $E(\sum _i p_i \hat{\rho}^{AB}_i) \le \sum _i p_i E(\hat{\rho}^{AB}_i)$;\\
(v)\textit{Partial additivity}, $E(\hat{\rho}^{\otimes n})=n E(\hat{\rho})$;\\
(vi)\textit{Partial continuity}, if state $\hat{\rho}_n$ approaches ${\hat{\psi}}^{\otimes n}$ for large $n$: $\langle{{\psi}^{\otimes n}}|\hat{\rho}_n |{{\psi}^{\otimes n}}\rangle \to 1$ for $n \to \infty $, then $|E({\hat{\psi}}^{\otimes n})-E(\hat{\rho}_n)|/n \to 0$.\\
\\
It is shown that for any measure $E$ satisfying the above set of conditions,
the following inequality holds for any state $\hat{\rho}$ \cite{18}:
\begin{eqnarray}
{E_D}(\hat{\rho})\le E(\hat{\rho})\le {E_C}(\hat{\rho}),
\label{eq:edeef}
\end{eqnarray}
where ${E_D}(\hat{\rho})$ is the distillable entanglement \cite{21} and
${E_C}(\hat{\rho})$ is the entanglement cost \cite{19}. 
Using this relation,
we can easily see that 
$E(\hat{\rho}_3^{\rm{out}}) > E(\hat{\rho}_2^{\rm{out}})=0$ 
since $\hat{\rho}_3^{\rm{out}}$ is entangled and any two-qubit entangled
state is distillable \cite{14}, namely,
$0<{E_D}(\hat{\rho}_3^{\rm{out}})\le E(\hat{\rho}_3^{\rm{out}})$.
We can also prove $E(\hat{\rho}_3^{\rm{in}}) < E(\hat{\rho}_2^{\rm{in}})$
for $0<t<0.495\cdots$ as follows.
From Eq.~(\ref{eq:edeef}), we have
\begin{eqnarray}
E(\hat{\rho}_3^{\rm{in}})\le {E_C}(\hat{\rho}_3^{\rm{in}}),\ \ {E_D}(\hat{\rho}_2^{\rm{in}})\le E(\hat{\rho}_2^{\rm{in}}), 
\label{eq:eecede}
\end{eqnarray}
and hence what we need is an upper bound on 
${E_C}(\hat{\rho}_3^{\rm{in}})$
and a lower bound on 
${E_D}(\hat{\rho}_2^{\rm{in}})$.

The entanglement cost ${E_C}(\hat{\rho})$ is upper-bounded
\cite{19} by the entanglement of formation ${E_f}(\hat{\rho})$ 
\cite{20},
which can be computed through the concurrence $C(\hat{\rho})$ 
\cite{3} as
\begin{eqnarray}
{E_f}(\hat{\rho})=H[\frac{1}{2}(1+\sqrt{1-{C^2}(\hat{\rho})})],\\
\label{eq:eformation}
H(p)=-p \log p -(1-p) \log (1-p).
\label{eq:h}
\end{eqnarray}
Using this relation, we have 
\begin{eqnarray}
{E_C}(\hat{\rho}_3^{\rm{in}})&\le&
{E_f}(\hat{\rho}_3^{\rm{in}})
\label{eq:ecef} \\
&=&H[\frac{1}{2}(1+\sqrt{1-t^2})].
\label{eq:efrho3in}
\end{eqnarray}
A lower bound of $E_D$ is given \cite{20,21} as
\begin{eqnarray}
g(\hat{\rho})=1+\sum_x \lambda_x \log \lambda_x\le {E_D}(\hat{\rho}),
\label{eq:sed}
\end{eqnarray}
where $\lambda_x$ are the diagonal entries of the matrix form of
$\hat{\rho}$ on a Bell basis. 
Since $\hat{\rho}_2^{\rm{in}}$ is rewritten on a Bell basis as
\begin{eqnarray}
\displaystyle
\frac{1}{6}
\left(
\begin{array}{cccc}
5& 0& 0& 1 \\
0& 0& 0& 0 \\
0& 0& 0& 0 \\
1& 0& 0& 1 \\
\end{array} 
\right),
\label{eq:rho'2in}
\end{eqnarray}
$g(\hat{\rho}_2^{\rm{in}})=1-H(1/6)$ $(\approx 0.35)$.
Solving $1-H(1/6)>H[(1+\sqrt{1-t^2})/2]$ numerically,
we find that
\begin{eqnarray}
{E_f}(\hat{\rho}_3^{\rm{in}})<g(\hat{\rho}_2^{\rm{in}}),
\label{eq:efs}
\end{eqnarray}
holds for $0<t<0.495\cdots$. From Eqs.~(\ref{eq:eecede}), (\ref{eq:ecef}), (\ref{eq:sed}), and 
(\ref{eq:efs}), we obtain
\begin{eqnarray}
E(\hat{\rho}_3^{\rm{in}})< E(\hat{\rho}_2^{\rm{in}}).
\label{eq:maineq}
\end{eqnarray}
Hence for $0<t<0.495\cdots$, the strict inequality (\ref{eq:strict}) holds 
for the family of entanglement measures satisfying the properties (i$'$), (ii)-(vi).

\section{Conclusion}
\label{sec:conclu}

We have shown examples of a local qubit channel 
exhibiting the selective entanglement breaking and 
an example with the strong selective entanglement breaking.
These results imply that even for the system as small as 
a qubit, a quantum channel/operation can have a preference over 
which kind of entanglement to break. In our examples,
the ordering with respect to entanglement is determined by 
the transformability through LOCC operations, and hence 
is defined solely by the property of monotonicity. 
This makes our results independent of the choice of the entanglement measure.
We have also shown that the ordering change with a strict 
inequality holds for a family of measures 
satisfying a set of plausible conditions.

\begin{acknowledgments}
We thank Adam Miranowicz for helpful discussions.
This work was supported by 21st Century COE Program by the Japan Society
for the Promotion of Science and a MEXT Grant-in-Aid for Young
Scientists (B) No. 17740265.
\end{acknowledgments}

\end{document}